\documentclass[conference]{IEEEtran}
\IEEEoverridecommandlockouts
\usepackage{cite}
\usepackage{amsmath,amssymb,amsfonts}
\usepackage[dvipdfmx]{graphicx}
\usepackage{textcomp}
\usepackage{xcolor}
\usepackage{algpseudocode}

\usepackage[ruled,vlined]{algorithm2e}
\def\BibTeX{{\rm B\kern-.05em{\sc i\kern-.025em b}\kern-.08em
    T\kern-.1667em\lower.7ex\hbox{E}\kern-.125emX}}

\begin{document}

\newtheorem{proposition}{Proposition}

\title{{Band Assignment in Ultra-Narrowband (UNB) Systems for Massive IoT Access}\\ 
\thanks{The work of Petar Popovski has been in part supported by the Danish Council for Independent Research, Grant Nr. 8022-00284B SEMIOTIC.}
}

\author{\IEEEauthorblockN{Enes~Krijestorac, Zheang Huai, Ghaith Hattab, Petar Popovski, Danijela Cabric}
\IEEEauthorblockA{{Electrical and Computer Engineering Department,} {University of California, Los Angeles} \\ 
Los Angeles, USA\\
		enesk@ucla.edu,
		samerhanna@ucla.edu, danijela@ee.ucla.edu 
}
}
\author{
    \IEEEauthorblockN{Enes~Krijestorac\IEEEauthorrefmark{1}, Ghaith Hattab\IEEEauthorrefmark{1}, Petar Popovski\IEEEauthorrefmark{2}, Danijela Cabric\IEEEauthorrefmark{1}}
    \IEEEauthorblockA{\IEEEauthorrefmark{1}\textit{Electrical and Computer Engineering Department,} \textit{University of California, Los Angeles}, USA
    \\\{enesk, ghattab\}@ucla.com, danijela@ee.ucla.edu} 
    \IEEEauthorblockA{\IEEEauthorrefmark{2}\textit{Department of Electronic Systems}, \textit{Aalborg University}, Denmark
    \\\ petarp@es.aau.dk}
}

\maketitle

\begin{abstract}
In this work, we consider a novel type of Internet of Things (IoT) ultra-narrowband (UNB) network architecture that involves multiple multiplexing bands or channels for uplink transmission. An IoT device can randomly choose any of the multiplexing bands and transmit its packet. Due to hardware constraints, a base station (BS) is able to listen to only one multiplexing band. The hardware constraint is mainly due to the complexity of performing fast Fourier transform (FFT) at a very small sampling interval over the multiplexing bands in order to counter the uncertainty of IoT device frequency and synchronize onto transmissions. The objective is to find an assignment of BSs to multiplexing bands in order to maximize the packet decoding probability (PDP). We develop a learning-based algorithm based on a sub-optimal solution to PDP maximization. The simulation results show that our approach to band assignment achieves near-optimal performance in terms of PDP, while at the same time, significantly exceeding the performance of random assignment. We also develop a heuristic algorithm with no learning overhead based on the locations of the BSs that also outperforms random assignment and serves as a performance reference to our learning-based algorithm.
\end{abstract}


\section{Introduction}
\IEEEPARstart{T}{he} Internet of Things (IoT) has the potential to change the technological landscape and bring great economical and societal benefits. The success of IoT on a large scale depends on several key enabling technologies, one of which is wireless communication.
Low-power wide-area (LPWA) networks are a new paradigm of wireless networking that is expected to become one of the key drivers of massive IoT \cite{raza2017low}. 
Compared to the legacy technologies, like cellular and short-range wireless networks, LPWA networks offer many benefits including wide-area connectivity for low-power and low-data-rate devices and low capital expenditure thanks to the use of the unlicensed spectrum. 
To enable long-range connectivity, LPWA networks primarily use sub-1GHz bands due to their favorable propagation conditions. Ultra-narrowband (UNB) LPWA solutions apply the ultra-narrowband transmissions, which enable demodulation at very low received power. Furthermore, UNB LPWA networks normally rely on simple ALOHA-like access protocols, where IoT devices avoid associating and synchronizing with any UNB basestation (BS); in essence, IoT devices operate in a broadcast mode, transmitting their packets at arbitrary time and frequency. Normally, the BSs operate in a decentralized manner and packet decoding occurs at the BSs, as opposed to a central cloud server where the received symbols from all BSs in the network could be combined together. Therfore, the packet is successfully transmitted if any BS decodes any of the packet transmissions. 

Since the bandwidth of transmissions is extremely low (on the order of hundreds of Hz), slotted channel access becomes infeasible due to random frequency drift of the local oscillator that becomes comparable to the bandwidth of transmissions in commodity hardware. UNB networks solve this problem by allowing the devices to transmit in an unslotted manner, while the receiving BSs do the task of accurately syncing on to a signal in frequency. 
This is accomplished by performing fast Fourier transform (FFT) at a very small sampling interval over the entire bandwidth of the multiplexing band, which is the portion of the spectrum across which UNB transmissions occur. 
Naturally, a UNB network would benefit from a wider multiplexing band. 
However, since the complexity of the FFT scales with the bandwidth of the multiplexing band, the multiplexing band has a feasibility limit on its bandwidth.
One way to introduce more frequency diversity would be to use several multiplexing bands with each BS associated to one band and IoT devices transmitting freely across any band. 
This would keep the capital expenditure the same since the BSs would still use the same hardware, only tuned to different multiplexing bands, while the capacity of the network could potentially increase. Indeed, in \cite{hattab2018spectrum} it has been shown that the capacity does increase by applying this paradigm at no additional cost. However, in \cite{hattab2018spectrum}, the assignment of BSs to bands was not given special attention and was assumed to be random, leaving room for further enhancements.

While the frequency allocation problems were previously researched in the context of many different wireless technologies, to the best of our knowledge, no problem setting that has been investigated so far is the same as the scenario considered in this paper and, therefore, no applicable solution for our problem already exists in literature. For example, channel assignment in 802.11 WLANs was previously investigated (see\cite{ling2006joint} and references therein). 
However, in WLANs, the user equipment (UE) is associated with a single access point while for the use case of UNB networks, the transmissions are broadcasted to all BSs. Channel assignment problems also appear in cognitive radio networks \cite{ahmed2014channel}, however, in none of the works that we surveyed, there exists a similar infrastructure to that of UNB LPWA networks. 
Finally, frequency reuse has received a great deal of attention with respect to cellular networks \cite{uygungelen2011graph, narayanan2001static, chae2011radio}, but in these problems, the goal is usually to assign frequency bands such that BSs that are in the neighborhood of one another do not receive the same band. This approach would not apply to our problem as its primary goal is to maximize the frequency utilization, while we are mainly concerned with maximizing packet decoding probability (PDP).       

In this work, we look for the optimal assignment of BSs to bands with the goal of maximizing the PDP of uplink packets in a multiband UNB network. 
We cast the assignment based on this objective as an integer non-linear problem (INLP), which we then approximate to find a sub-optimal solution. Based on the sub-optimal solution, we propose a two-step algorithm for achieving optimal assignments of BSs to bands in a multiband UNB network. The first step consists of learning the parameters required to solve the suboptimal INLP, while the second step solves the suboptimal INLP. Simulation results show us that the proposed algorithm substantially exceeds the performance of random assignment and closely matches the performance of an optimal solution. We also develop a heuristic algorithm with a lower learning overhead based on the locations of the BSs that also outperforms random assignment and serves as a performance reference to our training-based algorithm.

\section{System Model}
We consider a randomly distributed spatial topology of BSs, IoT devices, and interfering devices. We denote the number of BSs present in the network as $B$ and the set of all BSs is $\mathcal{B}=\{1,...,B\}$. The BSs are indexed by $b\in \mathcal{B}$. We assume a multiband access, i.e., we assume there are $M$ multiplexing bands, each of bandwidth $W$. The multiplexing bands are indexed by $m \in \mathcal{M}=\{1,...,M \}$. 

\begin{figure}
    \centering
    \includegraphics[width=0.7\linewidth]{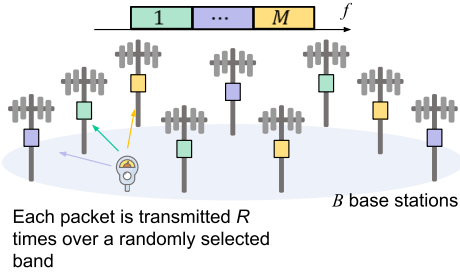}
    \vspace{-5pt}
    \caption{The system model of a multiband UNB network.}
    \label{fig:my_label}
    \vspace{-15pt}
\end{figure}

UNB IoT devices transmit signals at power $P_{IoT}$, occupying a bandwidth $w$. For the temporal generation of IoT traffic, UNB devices randomly transmit packets over time at a rate of $N$ packets per hour. Each packets is repeated $R$ times, consecutively over time, yet randomly hopping from one frequency to another, as shown in Fig. \ref{fig:my_label}. We label the transmissions with a tuple $n=[r,p]$, where $p$ is the packet index and $r=1,..,R$ corresponds to the repetition number of the packet. We denote the time of a transmission $n$ as $t(n)$. IoT devices are free to transmit on any frequency and on any of the multiplexing bands.
The signals are extremely narrowband and the channel access is assumed to be unslotted, therefore we model the carrier frequency $\phi(n)$ of the $n-$th transmission as a uniform random distribution $\mathcal{U}(w/2, MW - w/2)$. 
Similarly, since the devices randomly transmit across any of the multiplexing bands, we can model the band selected for a particular packet transmission $\beta(n)$ as a discrete uniform random variable that takes on the values $m \in \mathcal{M}=\{1,...,M \}$. The probability distributions across each band is $\mathbb{P}(\beta(n)=m)=\frac{1}{M}$.

Since the channel access is random, transmissions coming from different UNB devices in the network may interfere with one another. Two UNB transmissions, $n_1$ and $n_2$, will interfere if they overlap in frequency and time, i.e. $|\phi(n_1)-\phi(n_2)| \leq w$ and $|t(n_1)-t(n_2)| \leq T$. Furthermore, since the UNB network will likely exist in an unlicensed band, there will be additional interference due to devices using other technologies, such as LoRa or WiFi, or due to other coexisting UNB networks. Every interferer may not necessarily occupy the entire spectrum of $M$ multiplexing band and may instead only be causing interference across a portion of the bandwidth. In other words, the interference power is not necessarily independent of the multiplexing band.   

Our algorithm is designed to work in an environment where the channel gain can be modeled as a random process with three components: path loss, shadowing and fading. Let $P_b(n)$ denote the received power (in dBm) of a transmission $n$ from a source at location $q'_n$ to a BS $b$ at location $q_{b}$. 
$P_b(n)$ can then be expressed as $P_b(n)=P_{IoT} + P^{PL}_b(n)  + P^{SH}_b(n) + P^{F}(n)$, where $P^{T}$ is the transmitted power in dBm, $P^{PL}_b$ is the distance-dependent path-loss, $P^{SH}$ is a shadowing path loss and $P^{F}$ is the loss due to random fading. 
We consider a power-law path loss, $P^{PL}_b(n)=-10\alpha \log_{10}||q_b-q'_n||$, where $\alpha$ is the path-loss exponent. Fading gain, $P^{F}(n)$, is assumed to be an i.i.d. random variable when sampled over $n$. {We assume that $P^{SH}_b(n)$ is a spatially correlated random variable.} 


We denote the SINR of transmission $n$ at BS $b$ over multiplexing band $m$ as $\gamma_{b,m}(n)$, and it can be expressed (in dB) as 
$\gamma_{b,m}(n) = P_b(n) - (P^N(n) + P^I_{b,m}(n)),$ 
where $P^N(n)$ is the noise during transmission $n$ and $P^I_{b,m}(n)$ is the interference power during transmission $n$ at BS $b$ across multiplexing band $m$. $P^N(n)$ is an i.i.d. random variable and $P^I_{b,m}(n)$ is a random variable that is i.i.d. across $n$ but may be correlated across $b$ and $m$. For example, two nearby BSs or two neighboring multiplexing bands may experience interference from similar devices. 

A transmission $n$ is considered to be successfully decoded if the signal-to-noise-plus-interference ratio (SINR) at any of BS $b$ listening to the band $\beta(n)$ exceeds a threshold $\tau$. In practice, $\tau$ can be the minimum SINR required to achieve a certain bit error rate (BER) performance.
The value of $\tau$ depends on specific coding, modulation, and detection schemes being employed. For example, in the Sigfox technical documentation it is stated that an UNB transmission can be successfully decoded if its SINR exceeds 8 dB \cite{sigfox2017sigfox}. 

\title{Optimal band assignment for UNB LPWA networks with multiband access}

\section{Problem statement and a suboptimal solution}
In this section, we define the problem of optimal assignment and propose a suboptimal solution. The suboptimal solution is used to develop our proposed algorithm in Sec. \ref{sec:alg}.

\subsection{Problem statement}
We introduce a binary assignment variable $\mathbf{X}\in \mathbb{Z}_2^{B \times M}$, where $\mathbf{X}_{b,m}$ being equal to 1 indicates that BS $b$ is assigned to band $m$.
Our objective is to maximize the expected PDP of a packet $p$, given an assignment $\mathbf{X}$ and is given in the Eq. \ref{eq:ex_prob_1}. {We are interested in maximizing the probability of decoding of a packet irrespective of its source. The union over $r$ in Eq. \ref{eq:ex_prob_1} captures that if any of the repetitions $r=1,...,R$ is decoded, the packet $p$ is considered decoded. The union over $b$ captures that a repetition is decoded if any of the BSs $b$ is able to decode it. A BS can decode a repetition only if it is synchronized onto the band of the packet repetition $\beta([r,p])$ (i.e. $\mathbf{X}_{b,\beta([r,p])} = 1$) and if the SINR of the packet repetition at the basestation $b$, $\gamma_{b,\beta([r,p])}([r,p])$, exceeds the decoding threshold $\tau$. The decoding of individual repetitions of a packet will be correlated since they have the same source, therefore they may experience similar interference and shadowing. This is especially true when the repetitions happen to occur over the same multiplexing band. }

\begin{equation}
    P_{\text{decod,packet}}(\mathbf{X})=\mathbb{P}\left(\bigcup_{r}\bigcup_{b}\mathbf{X}_{b,\beta([r,p])}\gamma_{b,\beta([r,p])}([r,p])>\tau\right)
    \label{eq:ex_prob_1}
\end{equation}
The optimization problem can be stated as:
\begin{equation}
\begin{aligned}
\max_{\mathbf{X}}~ & P_{\text{decod,packet}}(\mathbf{X})\quad\textrm{s.t.}~ \mathbf{X}\mathbf{1}_M = \mathbf{1}_B,~\mathbf{X}\in \mathbb{Z}_2^{B \times M} 
\end{aligned}
\tag{P1}
\label{P1}
\end{equation}
where the constraints ensure one BS can only be assigned to one multiplexing band. In order to be able to solve this problem, we would have to know what the function  $\mathbb{P}\left(\bigcup_{r}\bigcup_{b}\mathbf{X}_{b,\beta([r,p])}\gamma_{b,\beta([r,p])}([r,p])>\tau\right)$ is in terms of the assignment variable $\mathbf{X}$. Learning or estimating this function would be non-trivial since the number of possible assignments is $M^B$. 
\subsection{Suboptimal solution to \ref{P1}}
To relax the optimization problem \ref{P1}, we seek to optimize the probability of decoding a {repetition of a packet, which is equivalent to maximizing the decoding probability of a transmission irrespective of the source},
$
    P_{\text{decod,trans.}}(\mathbf{X})=\mathbb{P}\left(\bigcup_{b}\mathbf{X}_{b,\beta([r,p])}\gamma_{b,\beta([r,p])}([r,p])>\tau\right)
    \label{eq:new_obj}
$, which is a lower bound on $P_{\text{decod,packet}}(\mathbf{X})$. 

The relaxed problem is then
\begin{equation}
\begin{aligned}
\max_{\mathbf{X}}~& P_{\text{decod,trans.}}(\mathbf{X})\quad \textrm{s.t.}~\mathbf{X}\mathbf{1}_M = \mathbf{1}_B,~\mathbf{X}\in \mathbb{Z}_2^{B \times M}
\end{aligned}
\tag{P2}
\label{P1.1}
\end{equation}

\begin{proposition}{}
For a given $\tau$, a suboptimal solution to \ref{P1.1} can be obtained by solving the following convex mixed-integer optimization problem:   
\begin{equation}
\begin{aligned}
\max_{\mathbf{X}} \quad & \sum_{m}\bigg(\sum_{b}X_{b,m}\mathbb{E}\{y_{b,m}(n)\}-\\
&\sum_{b<k}X_{b,m}X_{k,m}\mathbb{E}\{y_{b,m}(n)y_{k,m}(n)\}\bigg)\\
\textrm{s.t.} \quad & \mathbf{X}\mathbf{1}_M = \mathbf{1}_B,~\mathbf{X}\in \mathbb{Z}_2^{B \times M} \\
\end{aligned}
\tag{P3}
\label{P2}
\end{equation}
where $y_{b,m}(n)$ is Bernoulli random variable equal to 1, if transmission $n$ is decoded by BS $b$ while on band $m$, and 0 otherwise. The objective function is a concave function and is also a lower bound on the objective function in Eq. \ref{eq:ex_prob_1}.
\end{proposition}

\begin{IEEEproof}
We can express the objective function in (\ref{P1.1}), which is a lower bound on the objective function in (\ref{P1}), as $ P_{\text{det,trans.}}(\mathbf{X})=\sum_m\mathbb{P}(\beta(n)=m)\mathbb{P}\left(\bigcup_{b}\mathbf{X}_{b,m}\gamma_{b,m}(n)>\tau\right)$. 
Since we assume $\mathbb{P}(\beta(n)=m)$ is a constant across $m$, we can simply optimize over $P(\mathbf{X})=\sum_m\mathbb{P}\left(\bigcup_{b}\mathbf{X}_{b,m}\gamma_{b,m}(n)>\tau\right)$. Using exclusion-inclusion principle we can express this function as:
\begin{multline*}
    P(\mathbf{X})=\sum_{m}\Bigg(\sum_{b}\mathbb{P}\left(X_{b,m}\gamma_{b,m}(n)\geq\tau\right)-\\
    \sum_{b<k}\mathbb{P}\left(X_{b,m}\gamma_{b,m}(n)\geq\tau,X_{k,m}\gamma_{k,m}(n)\geq\tau\right)+\\
    ...+(-1)^{M-1}\sum_{b<...<k}\mathbb{P}\left(\bigcap_{b=1}^{K}X_{b,m}\gamma_{b,m}(n)\geq\tau\right)\Bigg)
\end{multline*}
We can take the entries of $\mathbf{X}$ out of the probability functions to get the expression:
\begin{multline}
    P(\mathbf{X})=\sum_{m}\Bigg(\sum_{b}X_{b,m}\mathbb{P}\left(\gamma_{b,m}(n)\geq\tau\right)-\\
    \sum_{b<k}X_{b,m}X_{k,m}\mathbb{P}\left(\gamma_{b,m}(n)\geq\tau,\gamma_{k,m}(n)\geq\tau\right)+\\
    ...+(-1)^{M-1}\sum_{b<...<k}\mathbb{P}\left(\bigcap_{b=1}^{K}X_{b,m}\gamma_{b,m}(n)\geq\tau\right)\Bigg)
    \label{eq:expansion}
\end{multline}
The above sum can be truncated up to the second-order terms
\begin{multline*}
P(\mathbf{X})\approx \Tilde{P}(\mathbf{X})=\sum_{m}\Bigg(\sum_{b}X_{b,m}\mathbb{P}\left(\gamma_{b,m}(n)\geq\tau\right)-\\
\sum_{b<k}X_{b,m}X_{k,m}\mathbb{P}\left(\gamma_{b,m}(n)\geq\tau,\gamma_{k,m}(n)\geq\tau\right) \Bigg)
\end{multline*}
The approximation $\Tilde{P}(\mathbf{X})$ is a concave function and is also a lower bound on the objective function. The latter follows from Bonferroni inequalities, since we are taking a second order approximation of the expression in Eq. \ref{eq:expansion} \cite{bonferroni1936teoria}. In $\Tilde{P}(\mathbf{X})$, $\mathbb{P}\left(\gamma_{b,m}(n)\geq\tau\right)$ is the expectation of $y_{b,m}$, $\mathbb{E}\{y_{b,m}\}$, and $\mathbb{P}\left(\gamma_{b,m}(n)\geq\tau,\gamma_{k,m}(n)\geq\tau\right)$ is the correlation $\mathbb{E}\{y_{b,m}y_{k,m}\}$. Hence, $\Tilde{P}(\mathbf{X})$ is the objective function in \ref{P2}.
\end{IEEEproof}
As shown by our results in the latter sections, approximating the problem (\ref{P1}) by (\ref{P2}) proves to be sufficient in finding the assignment that will result in the optimal decoding rate of transmissions in a realistic environment.
{While the solution given by (\ref{P2}) can be used to maximize the lower bound on packet decoding probability, in order to implement it, we need to know $\mathbb{E}\{y_{b,m}\}$ and $\mathbb{E}\{y_{b,m}y_{k,m}\}$. The implementation of this solution is considered in the next section.}

\section{The assignment algorithm}
\label{sec:alg}
Our algorithm is based on solving the suboptimal optimization problem (\ref{P2}), since this maximizes the lower bound on the expected probability of decoding a packet. Since the objective function in (\ref{P2}) is concave and the constraints are affine, this is a quadratic binary problem and can be efficiently solved using a quadratic integer programming solver. 

In order to solve for optimal assignment in a real setting, the estimates of $\mathbb{E}\{y_{b,m}(n)\}$ for $b \in \mathcal{B}$ and $m \in \mathcal{M}$, which we denote as $S_{b,m}=\hat{\mathbb{E}}\{y_{b,m}(n)\}$, and the estimates of $\mathbb{E}\{y_{b,m}(n)y_{k,m}(n)\}$ for $b,k \in \mathcal{B}$ and $m \in \mathcal{M}$, which we denote as $R_{b,k,m}=\hat{\mathbb{E}}\{y_{b,m}(n)y_{k,m}(n)\}$ are necessary. 
Naturally, to estimate these parameters training needs to be performed in the environment where the BSs are deployed. After these parameters are obtained or anytime they are updated over the course of the operation of the network, the problem \ref{P2} can be solved to obtain the optimal assignment. Therefore, our assignment algorithm can be divided into two stages:
\begin{itemize}
  \item \emph{Training stage}: The parameters $S_{b,m}$ and $R_{b,k,m}$ need to be estimated. In total, there are $KM$ parameters $S_{b,m}$  and $M{K \choose 2}$ parameters $R_{b,k,m}$. The estimation needs to be done periodically, as these parameters will be affected by the presence of the UNB devices in the network as well as presence of the interfering devices in the environment, both of which may change over time. {A central controller hosted at one of the BSs or in the cloud estimates these parameters based on the measurements collected by all BSs. }

  \item \emph{Optimization stage}: {The central controller solves the Problem (\ref{P2}) based on the estimates $S_{b,m}$ and $R_{b,k,m}$. }  

\end{itemize}

\subsection{The training stage}

In order to get the estimates of decoding rate $S_{b,m}$, the BSs need to be present on every band during training for at least some amount of time. 
Moreover, in order to get the estimate of the cross-correlation $R_{b,k,m}$ for all $b$, $k$ and $m$, every BS needs to experience every band simultaneously with every other BSs for at least some portion of the training. 



{We propose the following training approach. The training time is divided into $M$ slots, one for each multiplexing band. 
At each training slot, the BSs listen to a particular band and forward the captures of all detected but not necessarily decoded transmissions to a central processor. This, of course, means that IoT devices can transmit on only one band during a training slot, which reduces the network throughput during training. Nevertheless, we chose this approach since it is the fastest way to collect measurements. Other approaches to training that minimize the reduction in throughput during training could be devised, however this is outside the primary scope of this paper.} We assume that the central processor can identify all the unique transmissions that occurred from the captures $l \in \mathcal{L}$, where $\mathcal{L}=\{1,...,L\}$ and $L$ is the total number of unique transmissions detected during training. For the captures that were decoded by at least one BS, the transmissions can be identified based on the packet content, while for the minority of captures that have not been decoded by any BS but simply detected, this can be done based on the timestamps of their detection or based on IQ samples. The details of identification of unique transmissions are beyond the scope of this paper and for the rest of this paper we assume that the transmissions are identified correctly.  
Furthermore, the central processor records the multiplexing band that the BS is tuned to during transmission $l$ via variable $z_b(l)$, where $z_b(l)$ takes on the values in $\mathcal{M}$. For every transmission, $y_{b,m}(l)$ is also recorded.

After the training time is complete and all the transmissions are recorded, $S_{b,m}$ and $R_{b,k,m}$ can be obtained through 
\begin{equation}
S_{b,m} =\frac{1}{\sum_{l \in  \mathcal{L}}\mathbf{1}(z_b[l]=m)}\sum_{l \in  \mathcal{L}}y_{b,m}[l]    
\label{eq:s}
\end{equation}
\begin{equation}
R_{b,k,m} = \frac{1}{\sum_{l \in  \mathcal{L}}\mathbf{1}(z_b[l],z_k[l]=m)}\sum_{l \in\mathcal{L}}y_{b,m}[l]y_{k,m}[l]
\label{eq:r}
\end{equation}
The training procedure is summarized in Algorithm 1.

\begin{algorithm}[]
\DontPrintSemicolon
\SetAlgoLined
\KwResult{Estimates $S_{b,m}$, $R_{b,k,m}$}
$\mathcal{L} \leftarrow \emptyset$\;
\For{$m \in \mathcal{M}$}{
$\mathcal{L'} \leftarrow \emptyset$\;
Synchronize all BSs to band $m$\;
Collect all unique transmissions and add them to $ \mathcal{L'}$\;
    \For{$l \in \mathcal{L'}$}{
        $y_{b,m}[l] \leftarrow 0~\forall~m,b$\;
        $z_b[l]\leftarrow m~\forall~b$\;
        \For{$b \in \mathcal{B}$}{
            \If{$b$ decoded $l$}{
                $y_{b,m}[l] \leftarrow 1$}
        }
    }
$\mathcal{L} \leftarrow \mathcal{L} \cup \mathcal{L'}$\;
}
Perform updates in Eq. \ref{eq:s} and Eq. \ref{eq:r} 

\caption{Training procedure}
\end{algorithm}

While the proposed training procedure requires IoT devices not to transmit data, otherwise data loss may occur, we envision it to be employed less frequently. In between the long training procedures, more frequent updates can be made to subsets of parameters $S_{b,m}$ and $R_{b,k,m}$, requiring only a subset of BSs to move across bands.
\subsection{Low overhead training procedure}
\label{ref:low-overhead-training}
We can make further reduction in the overhead of our training procedure by assuming that $R_{b,k,m}$ is constant across multiplexing bands. In other words, we approximate $\gamma_{b,m}[n]$ to be identically distributed across $m$. {This would be appropriate if the interference $P^I_{b,m}(n)$ is independent of the multiplexing band $m$}. In this case, we would only have to learn $R_{b,k,m}$ on one band and apply it to all other bands. In total, we would have to learn ${K \choose 2}$ parameters $R_{b,k,m}$. The training procedure would have similar steps as the one shown in Algorithm 1, except measurements would only be collected on one band $m \in \mathcal{M}$.



\section{{Heuristic Algorithm based on BS Locations}}
\label{sec:simple_assign}
While our main contribution is the algorithm described in the previous section, we also propose a heuristic algorithm with a small training overhead that relies on the locations of the BSs to minimize the detection correlation between BS. This algorithm is based on the approximation that the detection correlation between two BSs $b$ and $k$ is dependent on the separation of the BSs as $\mathbb{E}\{y_{b,m}y_{k,m}\}\propto \left\Vert p_{b}-p_{k}\right\Vert ^{-\eta}$, where $\eta$ is a positive constant. This approximation is based on the observation that channel gain and interference of two BSs will be more similar the closer they are to each other. Furthermore, this approximation relies on detection correlation being independent on the multiplexing band which is the case only if the interference of the interfering users is uniformly distributed across all multiplexing bands.

To find the assignment via this heuristic approach we solve the problem
\begin{equation}
\min_{\mathbf{X}} ~ \sum_{b<k}X_{b,m}X_{k,m}C_{b,k} \quad
\textrm{s.t.}~ \mathbf{X}\mathbf{1}_M =\mathbf{1}_B,~\mathbf{X}\in \mathbb{Z}_2^{K \times M }
\tag{P4}
\label{P3}
\end{equation}
where
$$
C_{b,k}=\begin{cases}
1 & \text{if }b=k\\
\left\Vert p_{b}-p_{k}\right\Vert ^{-\eta} & \text{otherwise}
\end{cases}
$$
The parameter $\eta$ can be found empirically through trial and error. {The central controller can explore different values of $\eta$ and use the one that gives the best PDP after the assignment based on this heuristic algorithm is made.}
\section{Simulation results}

\subsection{Simulation environment}
We simulate an area of size $13 \text{x} 13~km^2$ with BSs, UNB devices and {interfering} devices randomly distributed over the area. At each Monte Carlo realization, the BS's locations are sampled from a uniform distribution. The UNB devices are generated from a homogeneous Poisson Point process (HPPP) $\Phi_{IoT}$ with density $\lambda_{IoT}$, while the locations of interfering devices are generated from a homogeneous an HPPP $\Phi_{I}$ with density $\lambda_{I}$. We choose a large training time of 1 h to demonstrate the full benefit of our algorithm, although the algorithm can operate with a shorter training period if necessary.   

Fading, $P^{F}(n)$, is modeled by a Rayleigh distribution with a scale parameter $\sigma_{F}$. $P^{SH}_b(n)$ is modeled as a spatially correlated zero-mean Gaussian random variable. The spatial correlation is modeled as  $\mathbb{E}\{ P_{SH,dB}(q_b,q'_1)P_{SH,dB}(q_b,q'_2)  \}=\sigma^2_{SH}e^{-||q'_1-q'_2||/\beta_F}$, where $\sigma^2_{SH}$ is the shadowing power, $\beta_{SH}$ is the decorrelation distance.

Interfering devices transmit signals at power $P_{I}$ over a bandwidth $w' \gg w$, where this bandwidth is assumed to be overlapped with the spectrum used by the UNB network. The interfering network could be another IoT-based network, e.g., LoRa. Interference between two devices occurs if their signals have any overlap in time and frequency. The interfering devices transmit at random, similar to the UNB devices. The duration of transmission for interfering devices is $T'$ and the number of packets per hour they transmit is $N'$. In order to introduce frequency diversity in interference, the interfering devices transmit only within a particular multiplexing band, unlike the UNB devices that randomly hop between multiplexing bands. The distribution of interfering devices across multiplexing bands is non-uniform in each simulation, therefore the levels of interference across each band are not even. Let $\beta'[i]$ be the multiplexing band of interfering device $i \in \mathcal{I}$, where $\mathcal{I}$ is the set of all interferers experienced by the UNB network. The probability that the interferer $i$ is assigned to operate in overlap with multiplexing band $b$ during simulation (i.e. the center frequency of its transmission is at a random frequency within the multiplexing band), $\mathbb{P}(\beta'[i]=m)$, is drawn from a uniform distribution $\mathcal{U}(0,1)$, subject to $\sum_{m \in \mathcal{M}}\mathbb{P}(\beta'[i]=m)=1$. This creates non-uniform levels of interference across each multiplexing band, which is what a real UNB network would experience.

\subsection{Results}
\begin{table}[t]
\vspace{0pt}
\renewcommand{\arraystretch}{1.3}
\caption{Parameter values used in simulation}
\label{table:params}
\centering

\begin{tabular}{c||c||c||c}
\hline 
Parameter & Value & Parameter & Value\tabularnewline
\hline 
\hline 
Noise power & $-146$ dBm & $w' $ & 125 kHz\tabularnewline
\hline 
$P_{IoT}$ & 14 dBm & $N'$ & $30$\tabularnewline
\hline 
$P_{I}$ & 14 dBm & Area size & 169 km$^2$ \tabularnewline
\hline 
$R$ & 3 & $B$  & 6\tabularnewline
\hline 
$N$ & $3$  & $\lambda_{IoT}$ & 5e3 m$^{-3}$ \tabularnewline
\hline 
Packet size & $2080$ b & $\lambda_{I}$  & 2e3 m$^{-3}$  \tabularnewline
\hline 
$w$ & $600$ Hz& $\tau$  & $10$ dB\tabularnewline
\hline 
Transmission time & $2080~b/w$ s & Training time  & 1 h \tabularnewline
\hline 
$W$ & $200$ kHz & $M$  & $3$ \tabularnewline
\hline 
$\sigma_{SH}$ & 9 dB & $\sigma_{F}$  & $1$ dB \tabularnewline
\hline 
\end{tabular}

\end{table}

Unless otherwise stated, the parameter values used during the simulation are specified in Table  \ref{table:params}. The proposed algorithm is compared against four benchmarks: 
\begin{enumerate}
    \item Random assignment
    \item The proposed algorithm using a low-overhead training procedure described in Section \ref{ref:low-overhead-training}
    \item The heuristic assignment based on BS locations as proposed in in Sec. \ref{sec:simple_assign}. We set $\eta$ to 1 since this maximizes the performance in our simulation environment.
    \item Optimal assignment that maximizes the probability of decoding a transmission $P_{\text{decod,trans.}}$
    \item Optimal assignment that maximizes the probability of decoding a packet, $P_{\text{decod,packet}}$
\end{enumerate} 
The theoretical best performance with optimal assignment for maximum packet decoding rate or maximum transmission decoding rate can easily be obtained in simulation. We rerun every Monte Carlo realization, with its particular sequence of pseudo-random events, for all possible $M^B$ assignments, and select the assignment that gives the highest average packet or transmission decoding rate. 

\begin{figure}[h!]
	\centering
	\vspace{-5pt}
	\includegraphics[width=0.95\linewidth]{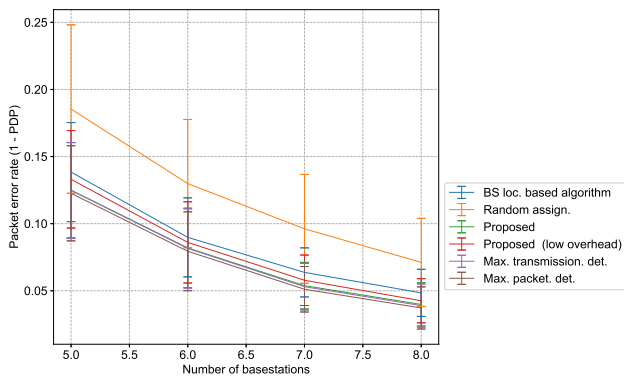}
	\vspace{-10pt}
	\caption{The performance of the proposed algorithm compared to the benchmarks for different number of BSs present in the network.}
	\label{fig:n_bs}

\end{figure}

First, we evaluate the performance of our algorithm for different numbers of BSs in the network while keeping all the other parameters constant. The results are shown in Fig. \ref{fig:n_bs}, where we measure the packet decoding error rate (1-PDP).  We can observe that the optimal assignment for maximum transmission decoding rate is close to optimal assignment for maximum packet decoding rate. This implies that the assignment problem  \ref{P1.1} is nearly equivalent to the assignment problem \ref{P1}. The error rate of our proposed algorithm matches the error rate that can be achieved by maximizing the transmission decoding probability. Using a training procedure with lower overhead impacts the performance of our algorithm since the interference characteristics are dependent on  the multiplexing band but this approach still outperforms BS location based heuristic algorithm. We can expect that the BS location based heuristic would perform even worse in an environment with a stronger shadowing and interference. While all the assignment algorithms outperform random assignment, random assignment also introduces higher variance in error rate which further motivates optimizing for assignment.   

\begin{figure}[h!]
	\centering
	\vspace{-5pt}
	\includegraphics[width=0.95\linewidth]{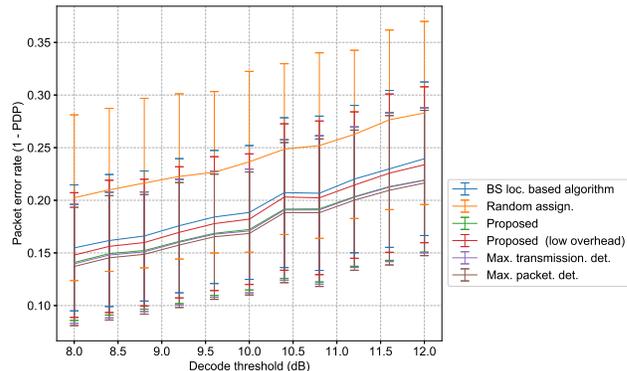}
	\vspace{-12pt}
	\caption{The performance of the proposed algorithm compared to the benchmarks for different decoding SINR thresholds in dB.}
	\label{fig:density}
    \vspace{-12pt}
\end{figure}

Next, we analyze the performance of our algorithm for different SINR decoding thresholds. The results are shown in Fig. \ref{fig:density}. The performance of each algorithm relative to each other remains the same. However, this graph further highlights the benefit of our proposed algorithm. For example, the proposed algorithm and the random algorithm achieve the same error rate at approximately 11 dB and 8 dB decoding thresholds, respectively. This means that with optimal assignment we can support a 3 dB higher decoding thresholds which translates to being able to use higher modulation order or higher coding rate.  

\section{Conclusions}
We have developed an algorithm for assignment of BSs to multiplexing bands in an UNB network in order to maximize the packet decoding probability. We find a theoretical suboptimal solution to this optimization problem and use this insight to develop an assignment algorithm that relies on the estimation of a small number of parameters. Furthermore, we introduce a heuristic algorithm that relies only on BS locations to perform assignment. Our results show that with optimal assignment the BSs can be better utilized for packet reception compared to random assignment. Furthermore, we show that simply relying on BS locations for assignment does not result in an optimal performance in an environment with a non-line-of-sight channel and frequency dependent interference.

\bibliography{references}
\bibliographystyle{ieeetr}

\end{document}